\documentclass{article}

\usepackage[english]{babel}

\usepackage[letterpaper,top=2cm,bottom=2cm,left=3cm,right=3cm,marginparwidth=1.75cm]{geometry}

\usepackage{amsmath,amssymb,amsthm}
\usepackage{graphicx}
\usepackage[colorlinks=true, allcolors=blue]{hyperref}

\newtheorem{definition}{Definition}
\newtheorem{theorem}{Theorem}

\newtheorem{example}{Example}
\newtheorem{corollary}{Corollary}
\newtheorem{lemma}{Lemma}
\newtheorem{proposition}{Proposition}
\newtheorem{fact}{Fact}

\newcommand{\K}{\mathbb K}
\newcommand{\FO}{$\mathrm{FO}$}
\newcommand{\suppo}{\mathrm{Supp}}

\newcommand{\NNF}{\mathrm{FO(NNF)}}
\newcommand{\adom}{\mathrm{adom}}
\newcommand{\pdi}{\pi_{{\bf A}(\bI)}}
\newcommand{\pdij}{\pi_{{\bf A}(\bJ)}}
\newcommand{\supp}{\mathrm {supp}}
\newcommand{\bI}{{\bf I}}
\newcommand{\bJ}{{\bf J}}

\providecommand{\dotdiv}{
  \mathbin{
    \vphantom{+}
    \text{
      \mathsurround=0pt 
      \ooalign{
        \noalign{\kern-.35ex}
        \hidewidth$\smash{\cdot}$\hidewidth\cr 
        \noalign{\kern.35ex}
        $-$\cr 
      }%
    }%
  }%
}

\usepackage{tikz}
\usetikzlibrary{arrows,chains,matrix,positioning,scopes}
\def\BibTeX{{\rm B\kern-.05em{\sc i\kern-.025em b}\kern-.08em
    T\kern-.1667em\lower.7ex\hbox{E}\kern-.125emX}}

\newcommand{\butnot}{\mathrel{%
  \hspace{.1ex}
  \begin{tikzpicture}[baseline=-.57ex, line width=.125ex]
    \draw[-] (.2ex,0) --(1.5ex,0);
    \draw[-, line width=.01ex, fill=black]
             (1.35ex,0) -- (1.84ex, .48ex)
                       -- (1.91ex ,.418ex)
                       -- (1.55ex,   0ex)
                       -- (1.91ex ,-.418ex)
                       -- (1.84ex,-.48ex)
                       -- (1.35ex,0ex);
  \end{tikzpicture}
\hspace{.1ex}}}

\usepackage{color,dsfont}

\title{Limitations of  Interpretations in Semiring Semantics}
\author{
Guillermo Badia\\University of Queensland\\ g.badia@uq.edu.au \and
Phokion G.\ Kolaitis\\University of California Santa Cruz\\
kolaitis@ucsc.edu}

\begin{document}
\maketitle

\begin{abstract}
Interpretations have been widely used in the study of provenance of queries on semiring-annotated databases. Here, we consider  two different interpretation-based semantics for first-order formulas in negation normal form, namely, the canonical-interpretation semantics and the compatible-interpretation semantics. We show that the resulting query languages cannot express the monus operation on semiring-annotated relations, when the annotations come from a positive semiring with monus and with at least three elements in its universe.  Consequently, the semantics of SQL cannot be captured using the interpretation-based semantics considered here. 

\subparagraph*{AI Statement:} No AI tools  were used in this research or in the writing of  this paper.

\end{abstract}

\section{Introduction}
During the past two decades, there has been a growing body of research on  semiring-annotated databases, where the tuples in the relations of a database are annotated with values from some semiring of interest.  This approach provides a unifying framework, known as semiring semantics, that contains as special cases both standard relational databases (as the databases over the Boolean semiring
${\mathbb B}=(\{0,1\}, \vee, \wedge, 0, 1\}$) and bag databases (as the databases over the bag semiring ${\mathbb N}=(N,+,\cdot,0,1)$). The catalyst for this investigation was the paper by Green, Karvounarakis, and Tannen \cite{GKV}, who used semirings of polynomials to formalize and study the provenance of database queries. The main focus
of \cite{GKV}  was on the provenance of unions of conjunctive queries for which intuitive and unambiguous semiring semantics can be given.  In the same paper,  semiring semantics were also given to Datalog programs (see also \cite{DBLP:conf/icdt/DeutchMRT14});   later on, semiring semantics were given to richer formalisms supporting recursion, such as  least-fixed point logic
\cite{DBLP:conf/csl/DannertGNT21}.

Conceptual difficulties arise when one attempts to develop semiring semantics for formalisms  that support  negation,  the reason being that several different approaches to the semantics of negation are possible. On the side  of relational algebra, for example, the question becomes: what is the ``right'' semiring semantics of the difference operation on annotated relations? To this effect, Geerts and Poggi \cite{GP}  gave semantics to the difference operation on annotated relations using the monus operation on semirings, while
Suciu \cite{DBLP:conf/birthday/Suciu24} considered two additional  approaches to giving semiring semantics to the difference operation. Note that
the monus operation as the semantics of difference is also adopted in the recent system ProvSQL~\cite{DBLP:journals/corr/abs-2609-31246, SenManiuSenellart2026ProvSQL}, which supports aggregate queries on semiring-annotated databases.  
 On the side of first-order logic (or relational calculus), the  connective $\butnot$ (`but not') was used in \cite{DBLP:journals/pacmmod/BadiaKN25}  to give semantics to first-order formulas. This connective encapsulates a limited form of negation and turns out to be the exact counterpart of the monus operation. Specifically, in \cite{DBLP:journals/pacmmod/BadiaKN25}, it was shown that,  relational algebra with the monus as the semantics of difference can express the same  queries on semiring-annotated databases  as first-order logic with the $\butnot$ (`but not') connective  does. This result not only generalizes
Codd's classical theorem about relational algebra and relational calculus \cite{Codd}, but also  yields an analogous theorem for bag databases, since  SQL uses the monus operation on the natural numbers  to give semantics to the difference operation on bags.

Gr\"adel and Tannen \cite{gradel2017semiring,G-T3} gave semantics to first-order formulas in negation normal form using interpretations.  By definition, an interpretation on a set $A$ is a function that assigns a semiring value to every atom or negated atom involving elements of $A$. Every fixed interpretation gives rise to semiring semantics of first-order formulas in negation normal form by lifting the interpretation from atoms and negated atoms to arbitrary first-order formulas in negation normal form. While interpretations have been successful in formalizing and studying the provenance of first-order formulas in negation normal form, they have not been used  to design languages for expressing queries over semiring-annotated databases so far, the main reason being that \cite{gradel2017semiring,G-T3}
 do not give interpretation-based semantics to first-order formulas in negation normal form  on semiring-annotated databases.

In this paper, we first address the following conceptual question: how can  interpretation-based semantics be used to define languages for expressing queries over semiring-annotated databases? We answer this question by introducing the notions of canonical-interpretation semantics and compatible-interpretation semantics. Both these notions give semantics to first-order formulas in negation normal form on semiring-annotated databases. Canonical-interpretation semantics is a stricter notion than compatible-interpretation semantics, because a special interpretation, called the canonical interpretation is associated with every semiring-annotated database, while in compatible-interpretation semantics  different interpretations can be used on different databases as long as they are ``compatible'' with the semiring-annotated database at hand, i.e., as long as they  agree with the annotations of the relations in the database under consideration.  

The main technical result reported here asserts that if $\mathbb K$ is a positive semiring with monus and with at least three elements in its universe, then the monus operation on $\mathbb K$-relations cannot be expressed by any first-order formula in negation normal form under the interpretation-based semantics introduced here.  In other words, the Boolean semiring is the only positive semiring $\mathbb K$ with monus for which the monus operation on $\mathbb K$-relations is expressible using interpretation-based semantics.
An immediate consequence of this result is that
the semantics of SQL cannot be captured using
canonical-interpretation semantics or compatible-interpretation semantics.

\section{Background}
This section contains a minimum amount of the necessary background material.

\subsection{Semirings with monus} \label{sec:semirings}
Here, we define the notion of a semiring with monus and give several examples.

A \emph{semiring} is a  structure $\mathbb{K}=(K, +, \cdot,0,1)$, where  addition $+$ and multiplication $\cdot$ are binary operations on $K$, and  $0$, $1$ are elements of $K$ such that
\begin{itemize}
\item $(K, +, 0)$  is a commutative monoid,  $(K, \cdot, 1)$ is a monoid, and $0\neq 1$;
\item 
for all   $a,b,c$ in $K$, we have that 
$a\cdot (b+c) = a\cdot b + a \cdot c$ and $(b+c) \cdot a = b\cdot a + c \cdot a$;
\item 
for all  $a$ in $K$, we have that $a \cdot 0 = 0 \cdot a = 0$. 
\end{itemize}

\noindent Let $\mathbb{K}=(K, +, \cdot,0,1)$ be a semiring. 

\begin{itemize}

\item  $\mathbb{K}$ is \emph{commutative}  if the monoid $(K,\cdot,1)$ is commutative.

\item $\mathbb{K}$ is \emph{zero-sum-free} if for all elements $a, b \in K$ with $a+b=0$, we have that $a=0$ and $b=0$.
\item 
$\mathbb{K}$ has
\emph{no zero divisors} if for all elements $a, b \in K$ with  $a\cdot b=0$, we have that $a=0$ or $b=0$.
\item  $\mathbb{K}$ is a  \emph{positive} semiring if it is commutative,  zero-sum-free, and has no zero divisors.

\item
The 
\emph{natural preorder} of $\mathbb{K}$ is the binary relation $\preceq$ on $K$, where   $a\preceq b$ holds if and only if there is some  $c\in K$ such that $a+c=b$. 
Note that  $\preceq$ is indeed a preorder on $K$, since it  is clearly a reflexive and transitive relation on $K$.
\item 
We say that $\preceq$
is the \emph{natural order} of $\mathbb K$ if $\preceq$ is a partial order on $K $, i.e., if   $a\preceq b$ and $b\preceq a$, then  $a=b$.  In this case, we also say that $\mathbb{K}$ is a \emph{naturally ordered}
semiring.
\end{itemize}

\begin{fact} If a semiring $\mathbb K$
 is naturally ordered, then $\mathbb K$ is zero-sum-free.
 \end{fact}
 Indeed,  if $a+b=0$, then we have that $a\preceq 0$; furthermore, since $0+a=a$, we also have that  $0\preceq a$; thus,   $a=0$ because $\preceq$ is a partial order.  A similar argument yields $b = 0$.

Some, but not all, naturally ordered semirings can be expanded with a \emph{monus} operation that shares some of the desirable properties of the \emph{difference} operation on rings.

 A  \emph{semiring with monus} is an algebraic structure  $\mathbb{K}= (K, +, \cdot, -,  0, 1)$ such that:
\begin{enumerate}
\item  $(K, +, \cdot, 0, 1)$ is a naturally ordered commutative semiring;
\item  For all elements $a,b \in K$,  there is a smallest (in the sense of the natural order $\preceq$) element $c$ in $K$, denoted by $a-b$,  such that $a\preceq b+c$.
\end{enumerate}

The next proposition 
asserts that the monus operation can be characterized in terms of a set of four rather simple identities  \cite{amer1984equationally}; a self-contained proof of this result can be found in \cite{DBLP:conf/birthday/Suciu24}.

\begin{proposition}\label{monusprop}
Let\/ $\mathbb{K}= (K, +, \cdot,  0, 1)$ be a naturally ordered semiring and let $-'$ be a binary operation on $K$. Then, the following statements are equivalent:
\begin{enumerate}
\item  For all $a,b \in K$, we have $a -' b = a-b$, i.e., 
$a -' b$ is the smallest element c such that
$a\preceq b+c$.
\item The following four identities hold for  $-'$:
$$a -' a = 0;\quad  0 -' a = 0;\quad  a + (b -' a) = b + (a -' b);\quad a -' (b + c) = (a -' b) -' c.$$
\end{enumerate}

\end{proposition}

We now give some examples of extensively studied positive  semirings with monus.

\begin{example}\label{bool} {\rm The two-element Boolean semiring $\mathbb{B}= (\{\top, \bot\}, \wedge, \vee, \bot, \top)$ can be expanded to a semiring with monus in which $a -b=a\wedge \overline{b}$, where $\overline{b}$ is the  Boolean complement of $b$.}
\end{example}

\begin{example}{\rm 
The semiring $\mathbb{N}=  (N, +, \cdot,  0, 1)$ of the non-negative numbers, also known as the \emph{bag} semiring, is a naturally ordered semiring in which the natural order coincides with the standard order $\preceq$ on $N$. The bag semiring 
can be expanded to a semiring with monus in which
the monus operation is the \emph{truncated subtraction}  $x \dotdiv y$, that is,  $x \dotdiv y= x -y$ if $x\geq y$, and $x \dotdiv y =0$ if $x<y$.} 
\end{example}

\begin{example} {\rm The \emph{tropical} semiring $\mathbb{T} = (\mathbb{R}\cup\{\infty\},\min, +, \infty, 0)$ of the real numbers is a naturally ordered semiring in which the natural order is the reverse standard order on the set $\mathbb R$ of all real numbers, i.e., $x\preceq y$ if and only if $x\geq y$.
It can be expanded to a semiring with monus by setting
$x-y = x$ if $x < y$, and  $x-y =  \infty$ if $x\geq y$, where $<$ is  the standard order of the reals.} 
\end{example}

\begin{example}{\rm 
 The \emph{fuzzy} semiring $\mathbb{F}=([0,1], \max, \min,0,1)$ is a naturally ordered semiring in which the natural order coincides with the standard order $\preceq$ on the interval $[0,1]$. It can be  expanded to a semiring with monus by setting  $x-y = x$ if $x > y$, and 
$x-y = 0$ if $x\preceq y$.}
\end{example}

\begin{example} \label{exam:suciu} {\rm The \emph{provenance} semiring $\mathbb{N}[X]$ is  the  semiring of all polynomials with variables from  a set $X=\{x_1, \dots, x_n, \ldots \}$   and with coefficients from the set $N$ of non-negative integers; the operations of ${\mathbb N}[X]$ are the standard addition and multiplication of polynomials. 

The provenance semiring ${\mathbb N}[X]$ can be expanded to a semiring with monus as follows. 
For every tuple $\alpha=(\alpha_1,\ldots,\alpha_n) \in N^n$, let $x^\alpha$ be the monomial
 $x_1^{\alpha_1} x_2^{\alpha_2} \dots x_n^{\alpha_n}$, where we set $x_i^0=1$.
 For a finite set $I\subseteq \mathbb{N}^n$, let $f[X] = \sum_{\alpha \in I}f_\alpha x^\alpha$ and  $g[X] = \sum_{\alpha \in I}g_\alpha x^\alpha$ be  two polynomials in $\mathbb{N}[X]$. The monus operation is defined by letting $f[X] - g[X] = \sum_{\alpha \in I} (f_\alpha \dotdiv g_\alpha)x^\alpha$, where $\dotdiv$ is the truncated difference on $N$, that is, $f_\alpha\dotdiv g_\alpha= f_\alpha - g_\alpha$ if $f_\alpha \geq g_\alpha$, and $f_\alpha\dotdiv g_\alpha=0$ if $f_\alpha < g_\alpha$.

The semiring ${\mathbb N}[X]$ is called the provenance semiring due to its use in the study of database provenance, as initiated in \cite{GKV}.}

\end{example}

It should be pointed out that there are  naturally ordered semirings that cannot be expanded to a semiring with monus; for a concrete example,  see \cite{Monet2016}. 

\subsection{Semiring-annotated relations, structures, and databases}

Let $D$ be a fixed denumerable set, which  will serve as  the \emph{domain} of elements occurring in databases. For  $n\geq 1$, we write $D^n$ to denote the set of all $n$-tuples from $D$.


 Let  $\mathbb{K}=(K, +, \cdot, -, 0,1)$ be a positive semiring with monus and let $n\geq 1$ be a non-negative integer.
\begin{itemize}
\item  A $n$-ary $\mathbb{K}$-relation $R$ is a function $R \colon D^n\longrightarrow K$ of finite \emph{support}, that is to say, the set $\supp(R) := \{(d_1,\ldots,d_n)\in D^n: R(d_1,\ldots,d_n) \neq 0\}$ is finite.  We call $n$ the \emph{arity} of $R$.
\item  A \emph{$\mathbb{K}$-relation} is a $n$-ary $\mathbb{K}$-relation $R$, for some $n\geq 1$.

\item   If  $A\subseteq D$, then a \emph{$n$-ary $\mathbb{K}$-relation on $A$} is a $n$-ary $\mathbb{K}$-relation $R$ with $\supp(R)\subseteq A^n$.
\end{itemize}

For example, if $\mathbb B$ is the Boolean semiring, then the $\mathbb B$-relations are the standard relations in databases.
Similarly, if $\mathbb N$  is the bag semiring, then the $\mathbb N$-relations are the bags (or multisets), i.e., relations in which each tuple has a non-negative integer as its multiplicity.

 A \emph{relational schema}  is a finite sequence $\tau= (\hat{R_1}, \ldots, \hat{R_p})$ of relation symbols, where each relation symbol $\hat{R_i}$ has a designated positive integer $r_i$ as its \emph{arity}.
We now consider the notions of a (finite) $\mathbb K$-structure and a $\mathbb K$-database over a relational schema $\tau$.

\begin{itemize}

\item  A \emph{$\mathbb{K}$-structure over $\tau$} is a tuple $\mathbf{A}= (A,R_1, \dots, R_p)$,
where $A$ is a non-empty finite subset of $D$, called the \emph{universe} of $\mathbf{A}$,
and each $R_i$ is a $\mathbb{K}$-relation on $A$
whose arity matches the arity  of the  relation symbol $\hat{R_i}$,
$1\preceq i\preceq p$.
\item A  \emph{$\mathbb{K}$-database over $\tau$} is a finite sequence $\mathbf{I}=(R_1,\ldots,R_p)$ of\/ $\mathbb{K}$-relations such that the arity of each $\mathbb{K}$-relation $R_i$ matches the arity of the  relation symbol $\hat{R_i}$, $1\preceq i\preceq p$.

\item Let $\mathbf{I}=(R_1,\ldots,R_p)$ be a $\mathbb{K}$-database. The \emph{active domain}  of $\mathbf{I}$, denoted by $\mathrm{adom}(\mathbf{I})$, is the set of all elements of $D$ that occur in the support of 
some relation $R_i$ of $\mathbf{I}$.

\item Every $\K$-database $\bI=(R_1^\bI,\ldots,R_m^\bI)$  gives rise to a  $\K$-structure ${\bf A}(\bI)$ with universe the active domain $\adom(\bI)$ of $I$ and with relations those of $\bI$, i.e., ${\bf A}(\bI)= (\adom(\bI),R_1^\bI,\ldots,R_m^\bI)$.

\end{itemize}

Clearly, $\mathbb B$-structures are finite relational structures, $\mathbb B$-databases are relational databases, while  $\mathbb N$-databases are bag databases.

Next, we define the fundamental notions of a query on $\mathbb K$-structures and a query on $\mathbb K$-databases.
\begin{itemize}
    
    \item Let $n\geq 1$.
An $n$-ary \emph{query $q$  on $\mathbb K$-structures}  is a mapping defined on $\mathbb{K}$-structures and such that if $\mathbf{A}=(A,R_1,\ldots,R_p)$ 
is a  $\mathbb{K}$-structures, then $q(\mathbf{A})$ is  a $n$-ary $\mathbb{K}$-relation   on $A$.
    \item
Let $n\geq 1$.
An $n$-ary \emph{query $q$  on $\mathbb K$-databases}  is a mapping defined on $\mathbb{K}$-databases and such that if $\mathbf{I}$ 
is a  $\mathbb{K}$-database, then $q(\mathbf{I})$ is  a $n$-ary $\mathbb{K}$-relation   on $\mathrm{adom}(\mathbf{I})$.
\end{itemize}
Clearly, a query $q$ on $\mathbb K$-structures gives rise to a query on $\mathbb K$-databases obtained by restricting $q$ to 
$\mathbb K$-structures of the form
${\bf A}(\bI)= (\adom(\bI),R_1^\bI,\ldots,R_m^\bI)$, where $\bI$ is a $\mathbb K$-database.

A \emph{query language} is a formalism for expressing queries. Codd \cite{DBLP:journals/cacm/Codd70,Codd} introduced relational algebra and relational calculus as formalisms for expressing queries on relational databases. Moreover, he showed that, in a precise sense, these two formalisms have the same expressive power. 
In \cite{DBLP:journals/pacmmod/BadiaKN25}, basic relational algebra 
$\mathcal{BRA}$ and basic relational calculus $\mathcal{BRC}$ were introduced as formalisms for expressing queries on $\mathbb K$-databases, where 
$\mathbb K$ is a positive semiring with monus. 

Basic relational algebra $\mathcal{BRA}$ is a procedural query language in which queries are specified by  expressions involving relations and operations on $\mathbb K$-relations. The syntax of $\mathcal{BRA}$-expressions over a relational schema $\tau=(\hat{R}_1,\ldots,\hat{R}_p)$ is given by the context-free grammar
$$E:= \hat{R}_i ~|~  E\cup E ~|~ E\cdot E ~|~ E \setminus E ~|~ \pi_v(E) ~|~ \sigma_\theta (E) ~|~ \supp(E).$$
The semantics of a $\mathcal{BRE}$-expression $E$ on a $\mathbb K$-database $\bf I$ is given by evaluating $E$ on $\bI$
and  obtaining a $\mathbb K$-relation $E^{\bf I}$ whose arity matches that of $E$. This evaluation uses operations on $\mathbb K$-relations (see~\cite[Def. 6]{DBLP:journals/pacmmod/BadiaKN25} for the precise definitions). For example, $\cup$,
$\cdot$, and $\setminus$ are interpreted using the union, the cartesian product, and the difference of two $\mathbb K$-relations, where these operations are defined as follows:
\begin{itemize}
\item 
If $R_1$ and $R_2$ are two $n$-ary $\mathbb{K}$-relations, then the \emph{union}  $R_1 \cup R_2$ is the $n$-ary $\mathbb{K}$-relation defined by
$(R_1 \cup R_2)({\bf t}) = R_1({\bf t}) + R_2({\bf t})$, for all ${\bf t} \in D^n$.
\item 
If $R_1$ is an $m$-ary $\mathbb{K}$-relation and  $R_2$ is an $n$-ary $\mathbb{K}$-relation, then the \emph{Cartesian product}  $R_1 \cdot R_2$ is the $(m+n)$-ary $\mathbb{K}$-relation defined by
$$(R_1 \cdot R_2)({\bf t}) = R_1(t_1,\ldots,t_m) \cdot R_2(t_{m+1},\ldots,t_{m+n}),$$ for all ${\bf t} =(t_1,\ldots,t_{m+n}) \in D^{m+n}$.

\item 
If $R_1$ and $R_2$ are two $n$-ary $\mathbb{K}$-relations, then the \emph{difference}  $R_1 \setminus R_2$ is the $n$-ary $\mathbb{K}$-relation defined by
$(R_1 \setminus R_2)({\bf t}) = R_1({\bf t}) - R_2({\bf t})$, for all ${\bf t} \in D^n$.

\end{itemize}

The basic relational calculus
$\mathcal{BRC}$ is a declarative query language in which queries are specified by formulas in a variant of first-order logic \FO~with  the connective 
$\butnot$ (`but not')  as a limited form of negation.\footnote{The  connective  $\butnot$ (`but not')  was also briefly considered in~\cite[\S 8.2]{G-T3}, but rejected 
 mainly based on the argument that it  does not give rise to a ``reasonable'' form of unrestricted negation.} The syntax of $\mathcal{BRC}$-formulas is given by the context-free grammar
$$\varphi:= x_i=x_j ~|~ \hat{R}_i(x_1,\ldots,x_{r_i}) ~|~ \varphi \vee \varphi ~|~ \varphi \land \varphi ~|~ \varphi \butnot \varphi ~|~ \exists x \varphi ~|~ \supp(\varphi).$$
As regards the semantics, with every $\mathcal{BRC}$-formula $\varphi$, every $\mathbb K$-structure ${\bf A}=(A,R_1,\ldots, R_p)$, and every assignment $\alpha$ mapping variables to elements of $A$, we associate a value   
 $\|\varphi\|(\mathbf{A}, \alpha)$ in $K$ that is defined recursively on the syntax of $\mathcal{BRC}$-formulas and simultaneously for all assignments as follows:
\begin{itemize}

\item $\|x_i=x_j\|(\mathbf{A},\alpha) =  1$,   if  $\alpha(x_i) = \alpha(x_j)$;~
$\|x_i=x_j\|(\mathbf{A},\alpha) =  0$, if $\alpha(x_i) \not = \alpha(x_j)$.

\item $ \|\hat{R_i} (x_{i_1},\ldots,x_{i_n})\|(\mathbf{A}, \alpha) = R_i(\alpha(x_{i_1}), \ldots \alpha(x_{i_n}))$, for $1\preceq i\preceq p$.

\item $\|\varphi_1 \vee \varphi_2\|(\mathbf{A},\alpha) = \|\varphi_1\|(\mathbf{A}, \alpha)+ \|\varphi_2\|(\mathbf{A}, \alpha)$; $\|\varphi_1 \wedge \varphi_2\|(\mathbf{A},\alpha) = \|\varphi_1\|(\mathbf{A}, \alpha) \cdot \| \varphi_2\|(\mathbf{A}, \alpha)$.
\item $|\varphi_1 \butnot  \varphi_2\|(\mathbf{A},\alpha) = \|\varphi_1\|(\mathbf{A}, \alpha) - \| \varphi_2\|(\mathbf{A}, \alpha)$
\item 
 $\|\supp (\varphi) \|(\mathbf{A},\alpha) = \mathsf{s} (\|\varphi\|(\mathbf{A}, \alpha)),  $
 where $s$ is the \emph{support} function on $K$ (i.e., for every $c\in K$, we have $s(c)=0$ if $s(c)\not =0$, and $s(c)=1$ if $s(c)=0$).
\item $\|\exists x\varphi\| (\mathbf{A}, \alpha) = \sum_{b\in A}\| \varphi\|(\mathbf{A}, \alpha[x/b])$, where $\alpha[x/b]$ is the assignment obtained from $\alpha$ by assigning the value $b$ to the variable $x$ while keeping everything else the same.
\end{itemize}
It is not hard to show that the value $||\varphi||({\bf A},\alpha)$ depends only on the values $a_1,\ldots,a_n$ of the assignment 
$\alpha$ on the  free variables, say $x_1,\ldots,x_n$, of $\varphi$. We write
$||\varphi||({\bf A},(a_1,\ldots,a_n))$ to denote
the value $||\varphi||({\bf A},\alpha)$, where $\alpha$ is any assignment such that $\alpha(x_1)=a_1,\ldots, \alpha(x_n)=a_n$.
Therefore, with every $\mathbb K$-structure ${\bf A}=(A,R_1,\ldots, R_p)$, we can associate a $n$-ary $\mathbb K$-relation $||\varphi||^{\bf A}$, where
$||\varphi||^{\bf A}(a_1,\ldots,a_n) = ||\varphi||({\bf A},(a_1,\ldots,a_n))$ if $(a_1\ldots,a_n)\in A^n$, while 
$\varphi^{\bf A}(a_1,\ldots,a_n) =0$ if $(a_1,\ldots,a_n)\not \in A^n$.  Note also that the syntax of $\mathcal{BRC}$ does not allow  universal quantification, but it does allow a limited form of negation via
the $\butnot$ connective.

In \cite{DBLP:journals/pacmmod/BadiaKN25}, the following version of
Codd's theorem for $\mathbb K$-relations was shown.

\begin{theorem}[\cite{DBLP:journals/pacmmod/BadiaKN25}] 
\label{Codd1} Let $\mathbb{K}=(K,+,\cdot, -,0,1)$ be a positive semiring with monus and  let $\tau=(\hat{R_1},\ldots \hat{R_p})$ be a relational schema. If  $q$ is  an $n$-ary query on $\mathbb K$-databases, then the following
statements are equivalent:
\begin{enumerate}
\item There is a $\mathcal{BRA}(\tau)$-expression $E$ of arity $n$ such that
$q(\mathbf{I})= E^\mathbf{I}$, for every $\mathbb{K}$-database  $\mathbf{I}$.
\item There is a $\mathcal{BRC}(\tau)$-formula
$\varphi(x_1,\ldots,x_n)$ such that $q(\mathbf{I})= ||\varphi||^{\mathbf{A}(\mathbf{I})}$, for every $\mathbb{K}$-database  $\mathbf{I}$, where ${\mathbf{A}}(\mathbf{I})$ is the $\mathbb K$-structure ${\mathbf{A}}(\mathbf{I}) = (\adom(\mathbf{I}), R_1,\ldots,R_p)$.
\end{enumerate}
\end{theorem}
Thus, basic relational algebra and basic relational calculus can, in a precise sense, express  the same queries on $\mathbb K$-databases, where $\mathbb K$ is a positive semiring with monus. In particular, this holds true for bag semantics, which is  the default semantics of SQL.

\subsection{Semiring semantics via interpretations}\label{A:truth}

Let $\NNF$ be the collection of all first-order formulas over $\tau$ in negation normal form.
 Gr\"adel and Tannen \cite{gradel2017semiring,G-T3} gave semiring semantics to $\NNF$-formulas  
 by using the notion of an \emph{interpretation}. 

In what follows,  we assume that ${\mathbb K}=(K, +, \cdot, 0,1)$ is a positive semiring,   $\tau = (\hat{R_1}, \ldots, \hat{R_p})$ is a relational schema, and $A$ is a finite subset of $D$.
\begin{itemize}
    \item  A \emph{literal from $A$}  is an atom or a  negated atom involving elements of $A$, i.e., it is  an  expression of the form $\hat{R}_i({\bf a})$ and $\neg \hat{R}_i({\bf a})$, where ${\bf a}$ is a tuple of elements from $A$. We write $\mathsf{Lit}_A$
to denote the set of all literals from $A$.
    \item An \emph{interpretation on $A$} is a function $\pi:\mathsf{Lit}_A \rightarrow K$.
    \end{itemize}
Let $\pi$ be an interpretation.
  Gr\"adel and Tannen \cite{gradel2017semiring,G-T3}, 
   associate a value $\pi[[\varphi]](\alpha)$ to every $\NNF$-formula $\varphi$  and every assignment $\alpha$ mapping variables to elements of $A$; this value is defined by induction on the definition of $\NNF$-formulas 
    and simultaneously for all assignments as follows:
  \begin{itemize}

\item $\pi[[x_i=x_j]](\alpha) =   1$,   if  $\alpha(x_i) = \alpha(x_j)$;~
$\pi[[x_i=x_j]](\alpha) =  0$, if $\alpha(x_i) \not = \alpha(x_j)$.
\item  $\pi[[x_i\not =x_j]](\alpha) =   1$,   if  $\alpha(x_i) \not = \alpha(x_j)$;~
$\pi[[x_i\not =x_j]](\alpha) =  0$, if $\alpha(x_i)  = \alpha(x_j)$.
\item $ \pi[[\hat{R_i} (x_{i_1},\ldots,x_{i_n})]](\alpha)  = \pi(\hat{R}_i(\alpha(x_{i_1}), \ldots \alpha(x_{i_n})))$, for $1\preceq i\preceq p$.

\item $ \pi[[\neg\hat{R_i} (x_{i_1},\ldots,x_{i_n})]](\alpha)  = \pi(\neg \hat{R}_i(\alpha(x_{i_1}), \ldots \alpha(x_{i_n})))$, for $1\preceq i\preceq p$.

\item $\pi[[\varphi_1 \vee \varphi_2]](\alpha) = \pi[[\varphi_1]](\alpha)+ \pi[[\varphi_2]](\alpha)$;~ $\pi[[\varphi_1 \wedge \varphi_2]](\alpha) = \pi[[\varphi_1]](\alpha) \cdot  \pi[\varphi_2]](\alpha)$.
\item $\pi[[\exists x\varphi]] (\alpha) = \sum_{b\in A}\pi[[ \varphi]](\alpha[x/b])$;~
 $\pi[[\forall x\varphi]] (\alpha) = \prod_{b\in A}\pi[[ \varphi]](\alpha[x/b])$
\end{itemize} 
Unlike the basic relational calculus, universal quantification is allowed here,  but negation is limited to negated atoms only.
A straightforward induction on the syntax of $\NNF$-formulas shows that the value
$\pi[[\varphi]](\alpha)$
 depends only on the values $a_1,\ldots,a_n$ of the assignment 
$\alpha$ on the  free variables, say $x_1,\ldots,x_n$, of $\varphi$. In what follows, we will write
$\pi[[\varphi]](x_1/a_1,\ldots,x_n/a_n)$ to denote
the value $\pi[[\varphi]](\alpha)$, where $\alpha$ is any assignment such that $\alpha(x_1)=a_1,\ldots, \alpha(x_n)=a_n$.  When the variables are understood from the context, we will simply write
$\pi[[\varphi]](a_1,\ldots,a_n)$.

Clearly,
if $\psi$ is a \FO-sentence in negation normal form and $\pi$ is an interpretation $\pi$,
then the value $\pi[[\psi]](\alpha)$ is the same for all assignments $\alpha$ mapping variables to elements of $A$.
We will write  $\pi[[\psi]]$ to denote this semiring value.

The following restricted notion
of an interpretation was considered in \cite{gradel2017semiring,G-T3}.
\begin{itemize}
 \item An interpretation $\pi$ is \emph{model defining} if for every atom $\hat{R}_i({\bf a})$, exactly one of the values $\pi(\hat{R}_i({\bf a}))$ and $\pi(\neg \hat{R}_i({\bf a}))$ is equal to $0$.
 \end{itemize} 
 The  intuition behind this notion is  that no two complementary literals are allowed to be  simultaneously \emph{false} (since $0$ is the smallest element  in a naturally-ordered semiring).

 Every  model-defining interpretation $\pi$  on $A$  gives rise to  a structure ${\bf A}_\pi=(A,R_1^{{\bf A}_\pi}, \ldots, R_p^{{\bf A}_\pi})$ with universe $A$, where for every $i\preceq p$ and for every tuple ${\bf a}$ from $A$, we have that
    ${\bf a}\in R_i^{{\bf A}_\pi}$ if and only if $\pi(R_i({\bf a}))\not = 0$.
The following proposition (Proposition 5 in \cite{G-T3}) will be useful in the sequel.

\begin{proposition} \label{pro:model-def}
Let $\pi$ be a model-defining interpretation on $A$. Then for every $\NNF$-sentence $\psi$, we have that
$$\pi[[\psi]] \not = 0 \quad \mbox{if and only if} \quad {\bf A}_\pi\models \psi.$$
\end{proposition}

\section{Interpretation-based query languages}
In this section, we address the following conceptual question:
how can  interpretation-based semantics be used to define languages for expressing queries over $\mathbb K$-databases?

It should be noted that Gr\"adel and Tannen \cite{gradel2017semiring,G-T3} do not give interpretation-based semantics to $\NNF$-formulas   on ${\mathbb K}$-structures.  In particular, they do not define what it means for a $\mathbb K$-structure $\bf A$   to \emph{satisfy} a \FO-sentence  $\psi$ in negation normal form. In \cite{DBLP:journals/pacmmod/KolaitisPVW25}, a definition of the latter concept was given in order to study repairs of $\mathbb K$-databases. This was done by introducing the notions of compatible interpretation and   canonical interpretation  on a $\mathbb K$-structure.

\begin{definition} \label{compatible-defn}
Let ${\bf A}=(A,R_1^{\bf A}, \ldots,R_p^{\bf A})$ be a  $\mathbb K$-structure over a relational schema $\tau$.

An interpretation $\pi$ on $A$ is \emph{compatible with} $\bf A$ if $\pi$ is model-defining and for every $i\preceq p$ and every tuple ${\bf a}$ with elements from $A$, we have that $\pi(\hat{R}_i({\bf a}))= R_i^{\bf A}({\bf a})$.

\end{definition}

Note that if $\pi$ is an  interpretation that is compatible with a $\mathbb K$-structure $\bf A$, then, by definition, $\pi$ has to agree with each relation $R_i^{\bf A}$ on the atomic facts $\hat{R}_i({\bf a})$. For the negated atomic facts $\neg \hat{R}_i({\bf a})$, we have that 
$\pi(\neg \hat{R}_i({\bf a})) = 0$ if $R_i^{\bf A}({\bf a})\not =  0$, because $\pi$ is model defining.  If, however,  $R_i^{\bf A}({\bf a}) =  0$, then $\pi(\neg \hat{R}_i({\bf a}))$ can be any non-zero value from $K$. We now introduce a special compatible interpretation.

\begin{definition}
Let ${\bf A}=(A,R_1^{\bf A}, \ldots,R_p^{\bf A})$ be a  $\mathbb K$-structure over a relational schema $\tau$.

 The \emph{canonical  interpretation on ${\bf A}$} is the interpretation $\pi_{\bf A}$ such that for every $i\preceq p$ and every tuple ${\bf a}$ of elements from $A$, we have that $\pi_{\bf A}(\hat{R}_i({\bf a}))= R_i^{\bf A}({\bf a})$ and
$$
\pi_{\bf A}(\neg \hat{R}_i({\bf a}))=\begin{cases}
			0, & \text{if $R_i^{\bf A}({\bf a })\not =0$}\\
            1, & \text{if $R_i^{\bf A}({\bf a}) =0$.}
		 \end{cases}
$$
\end{definition}
Clearly, the canonical interpretation $\pi_{\bf A}$  is compatible with $\bf A$; it is ``canonical'' in the sense that it assigns $1$ as the default value to every literal $\neg \hat{R}_i({\bf a})$ such that
$R_i^{\bf A}({\bf a})=0$.

Let $\bf A$ be a $\mathbb K$-structure and consider the canonical  interpretation $\pi_{\bf A}$ on $\bf A$.  Since $\pi_{\bf A}$ is model-defining,  it gives rise, as discussed earlier, to a  relational structure ${\bf A}_{\pi_{\bf A}}$ with universe $A$. From the definitions, it follows that
${\bf A}_{\pi_{\bf A}}= \suppo({\bf A})$, where $\suppo({\bf A})=(A, \suppo(R_1^{\bf A}), \ldots, \suppo(R_m^{\bf A}))$ is the  relational structure with universe $A$ and with  relations the supports of the $\mathbb K$-relations of $\mathbb A$.


We now introduce the notion of a query being definable using canonical-interpretation semantics.

\begin{definition} \label{defn:can-int} 
Let $n$ be a positive integer.
\begin{itemize}
\item 
A $n$-ary query $q$  on $\mathbb K$-structures is \emph{definable under canonical-interpretation semantics} if there is a $\NNF$-formula $\varphi(x_1,\ldots,x_n)$ such that for 
every ${\mathbb K}$-structure ${\bf A}= (A,R_1^{\bf A},\ldots,R_p^{\bf A})$ and every tuple $(a_1,\ldots,a_n)$, we have that
$$
q({\bf A})(a_1,\ldots, a_n) =
\begin{cases}
 \pi_{\bf A}[[\varphi]](a_1,\ldots,a_n), & \text{if} ~
(a_1,\ldots,a_n)\in A^n \\
0, & \text{if} ~ (a_1,\ldots,a_n)\not \in A^n,
\end{cases}
$$
where $\pi_{\bf A}$ is the canonical  interpretation on the $\mathbb K$-structure ${\bf A}$.
\item A $n$-ary query $q$  
on $\mathbb K$-databases is \emph{definable under canonical-interpretation semantics} if there is a $\NNF$-formula $\varphi(x_1,\ldots,x_n)$ such that for every $\mathbb K$-database ${\bf I}=(R_1^\bI,\ldots,R_p^\bI)$ and every tuple $(a_1,\ldots,a_n)$, we have that
$$q({\bf I})
(a_1,\ldots, a_n) = 
\begin{cases}
\pi_{{\bf A}(\bI)}[[\varphi]](a_1,\ldots,a_n), & \text{if} ~
(a_1,\ldots,a_n)\in \adom({\bf I})^n \\
0, & \text{if} ~ (a_1,\ldots,a_n)\not \in \adom({\bf I})^n,
\end{cases}
$$
where $\pi_{{\bf A}(\bI)}$ is the canonical  interpretation on the $\mathbb K$-structure ${\bf A}(\bI)=(\adom(\bI),R_1^\bI,\ldots,R_p^\bI)$.
\end{itemize}

\end{definition}

Next, we  introduce the notion of a
 query being definable under a much more relaxed notion of interpretation-based semantics.

 \begin{definition} \label{dfn:comp-int}
 Let $n$ be a positive integer.
 \begin{itemize}
 \item 
A $n$-ary query $q$  on $\mathbb K$-structures is \emph{definable under compatible-interpretation semantics} if
there is a $\NNF$-formula $\varphi(x_1,\ldots,x_n)$ such that for every $\mathbb K$-structure ${\bf A}=(A,R_1^{\bf A},\ldots,R_p^{\bf A})$, there is an interpretation $\pi$ that is compatible with $\bf A$ and such that 
for every tuple $(a_1,\ldots,a_n)$, we have that 
$$
q({\bf A})
(a_1,\ldots, a_n) = 
\begin{cases}
\pi[[\varphi]](a_1,\ldots,a_n), & \text{if} ~
(a_1,\ldots,a_n)\in A^n \\
0, & \text{if} ~ (a_1,\ldots,a_n)\not \in A^n.
\end{cases}
$$
\item
A $n$-ary query $q$   
on $\mathbb K$-databases is \emph{definable under compatible-interpretation semantics}
 if there is a $\NNF$-formula $\varphi(x_1,\ldots,x_n)$ such that for every $\mathbb K$-database ${\bf I}=(R_1^\bI,\ldots,R_p^\bI)$, there is an interpretation $\pi$ that is compatible with ${\bf A}({\bf I})$ and such that
 for every tuple $(a_1,\ldots,a_n)$, we have that 
 $$
q({\bf I})
(a_1,\ldots, a_n) = 
\begin{cases}
\pi[[\varphi]](a_1,\ldots,a_n), & \text{if} ~
(a_1,\ldots,a_n)\in \adom({\bf I})^n \\
0, & \text{if} ~ (a_1,\ldots,a_n)\not \in \adom({\bf I})^n.
\end{cases}
$$
\end{itemize}
 \end{definition}
Thus, canonical-interpretation semantics use a special compatible interpretation, while compatible-interpretation semantics do not commit to a particular compatible interpretation. Clearly, if a query is definable under the canonical-interpretation semantics, then it is also definable under the compatible-interpretation semantics. 
As we shall see next,  however, the converse need not be true.

Let $\tau$ be a relational schema consisting of a unary relation symbol $\hat{R}$.
Let ${\mathbb K}=(K,+,\cdot, 0, 1)$
be a positive semiring such that $K$ has at least three elements. Let $p$ and $q$ be two non-zero elements in $K$.  Since $p\preceq p+q$, we have that
$p\prec p+q$ or $p=p+q$. If $p=p+q$, then $q\prec p+q$, since, otherwise, we would have else $q=p+q=p$. So, we have that $p\prec p+q$ holds or $q\prec p+q$ holds.  So, let us  assume that $p \prec p+q$.
Let ${\bf A}_1$
and ${\bf A}_2$ be the following $\tau$-structures:
\begin{itemize}
\item ${\bf A}_1=(\{a,b\}, R^{{\bf A}_1})$, where
$R^{{\bf A}_1}(a)=0$ and $R^{{\bf A}_1}(b)=p$;
\item 
${\bf A}_2=(\{a,b\}, R^{{\bf A}_2})$, where
$R^{{\bf A}_2}(a)=0$ and $R^{{\bf A}_2}(b)=p+q$.
\end{itemize}
Furthermore, let $\pi_1$ and $\pi_2$ be the following interpretations on the set 
$\{a,b\}$:
\begin{itemize}
    \item $\pi_1(\hat{R}(a))=0$, 
    $\pi_1(\neg \hat{R}(a))=p+q$,
    $\pi_1(\hat{R}(b))=p$, 
    $\pi_1(\neg\hat{R}(b))=0$;
    \item 
$\pi_2(\hat{R}(a))=0$, 
    $\pi_2(\neg \hat{R}(a))=p$,
    $\pi_2(\hat{R}(b))=p+q$, 
    $\pi_2(\neg \hat{R}(b))=0$.
\end{itemize}
Clearly, $\pi_1$ is compatible with ${\bf A}_1$, while     $\pi_2$ is compatible with ${\bf A}_2$.

Let $q^*$ be the following unary query on $\tau$-structures:
\begin{itemize}
\item $q^*({\bf A}_1)=\pi_1[[\neg \hat{R}(x_1)]]$;
\item $q^*({\bf A}_2)=\pi_2[[\neg \hat{R}(x_1)]]$;
\item $q^*({\bf B})=\pi_{\bf B}[[\neg \hat{R}(x_1)]]$, if ${\bf B}$ is a $\tau$-structure that is different from both ${\bf A}_1$ and ${\bf A}_2$.
\end{itemize}
By construction, 
the
$\NNF$-formula
$\neg \hat{R}(x_1)$ defines
the query $q^*$ under compatible-interpretation semantics. Further, we have that $q^*({\bf A}_1)(a) = p+q$ and $q^*({\bf A}_2)(a) = p$, hence
$q^*({\bf A}_1)(a)\not \preceq
q^*({\bf A}_2)(a)$.

We now claim that no $\NNF$-formula $\varphi$ defines the query $q^*$ under canonical-interpretation semantics. 
Indeed, the  canonical interpretations $\pi_{{\bf A}_1}$ and $\pi_{{\bf A}_2}$ on  $\{a,b\}$ satisfy the pointwise comparison $\pi_{{\bf A}_1} \preceq \pi_{{\bf A}_2}$, as defined in  \cite[proof of Theorem 7 on p.\ 9]{DBLP:journals/corr/abs-1910-07910}. From the argument in the proof of \cite[Claim (2), Theorem 7 on p.\ 9]{DBLP:journals/corr/abs-1910-07910}, we get that $\pi_{{\bf A}_1}[[\varphi]](\alpha) \preceq \pi_{{\bf A}_2}[[\varphi]](\alpha)$ holds for every $\NNF$-formula and every assignment $\alpha$ that maps variables to elements in the set $\{a,b\}$ (the fully-chain complete condition on the semiring is only used for the least fixed-point operator).
In particular, we have 
$\pi_{{\bf A}_1}[[\varphi]](a)  \preceq \pi_{{\bf A}_2}[[\varphi]](a)$, while, as we saw earlier,
$q^*({\bf A}_1)(a)\not \preceq
q^*({\bf A}_2)(a)$. Thus, no $\NNF$-formula $\varphi$ can define the query $q^*$ under canonical-interpretation semantics. 

\smallskip

\section{Limitations of interpretation-based query languages} \label{sec:limitations}
We will show that the interpretation-based query languages introduced in the preceding section cannot express the monus operation on $\mathbb K$-databases.
As a stepping stone towards this goal, we introduce a relaxation 
of the notion of  compatible interpretation and prove a key technical  lemma. 

\begin{definition} \label{dfn:weak-comp}
Let ${\bf A}=(A,R_1^{\bf A}, \ldots,R_p^{\bf A})$ be a  $\mathbb K$-structure over a relational schema $\tau$.

An interpretation $\pi$ on $A$ is \emph{weakly compatible with} $\bf A$ if $\pi$ is model-defining and for every $i\preceq p$ and every tuple ${\bf a}$ with elements from $A$, we have that $\pi(\hat{R}_i({\bf a}))\not = 0$ if and only if $R_i^{\bf A}({\bf a})\not = 0$.
\end{definition}
Clearly, every compatible interpretation is also weakly compatible, but there are weakly compatible interpretations that are not compatible.

\begin{lemma} \label{lem:key} Let ${\mathbb K}=(K,+,\cdot,0,1)$ be a positive semiring with monus, 
let ${\bf A}=(A,R_1,\ldots,R_p)$ be a $\mathbb K$-structure, and let
$\pi_1$ and $\pi_2$ be two interpretations on $A$ that are weakly compatible with $\bf A$.
Then for every $\NNF$-formula $\varphi(x_1,\ldots,x_n)$ and  every assignment $\alpha$ mapping variables to values in $A$, we have that
$$\pi_1[[\varphi]](\alpha)\not = 0\quad\text{if and only if} \quad \pi_2[[\varphi]](\alpha) \not = 0.$$
\end{lemma}
\begin{proof}
The claim is proved by induction on the construction of $\NNF$-formulas and simultaneously for all assignments.
\begin{itemize}
\item Assume that $\varphi$ is a formula
of the form $x_i=x_j$ or of the form $x_i\not = x_j$.  Since $\pi_1$ and $\pi_2$ are interpretations, we have that  $\pi_1[[\varphi]](\alpha)= 
\pi_2[[\varphi]](\alpha)$, hence the result holds.
\item Assume that $\varphi$ is
an atom $\hat{R}(x_1,\ldots,x_r)$, where $\hat{R}$ is one of the relation symbols $\hat{R_1},\ldots,\hat{R_p}$.
By the definition of an interpretation, we have that $\pi_i[[\hat{R}(x_1,\ldots,x_r)]](\alpha) = \pi_i(\hat{R}(\alpha(x_1),\ldots, \alpha(x_r))$,
for $i=1,2$.
Since $\pi_1$ and $\pi_2$ are weakly compatible interpretations, we have that
$$\pi_1(\hat{R}(\alpha(x_1),\ldots, \alpha(x_r))\not = 0\quad\text{iff}\quad R^{\bf A}(\alpha(x_1),\ldots, \alpha(x_r)) \not = 0\quad\text{iff}\quad\pi_2(\hat{R}(\alpha(x_1),\ldots, \alpha(x_r))\not = 0.$$
Hence, we have that
$$\pi_1[[\hat{R}(x_1,\ldots,x_r)]](\alpha) \not = 0\quad\text{if and only if}\quad\pi_2[[\hat{R}(x_1,\ldots,x_r)]](\alpha)\not = 0.$$

\item Assume that $\varphi$ is
a negated atom $\neg \hat{R}(x_1,\ldots,x_r)$, where $\hat{R}$ is one of the relation symbols $\hat{R_1},\ldots,\hat{R_p}$.
By the definition of an interpretation, we have that $\pi_i[[\neg\hat{R}(x_1,\ldots,x_r)]](\alpha) = \pi_i(\neg\hat{R}(\alpha(x_1),\ldots, \alpha(x_r))$,
for $i=1,2$.
We now distinguish two cases:
\begin{enumerate}
\item If $R^{\bf A}(\alpha(x_1),\ldots,\alpha(x_r))=0$,
then, since $\pi_1$ and $\pi_2$ are weakly compatible interpretations, we have that 
$\pi_1(\hat{R}(\alpha(x_1),\ldots, \alpha(x_r)) = 0 = \pi_2(\hat{R}(\alpha(x_1),\ldots, \alpha(x_r)).$
Therefore, since $\pi_1$ and $\pi_2$ are model-defining interpretations, we have that
$\pi_i(\neg \hat{R}(\alpha(x_1),\ldots, \alpha(x_r)) \not = 0$, for $i=1,2$.  Consequently, we have that
$\pi_i[[\neg\hat{R}(x_1,\ldots,x_r)]](\alpha)\not = 0$, for $i=1, 2$.
\item If $R^{\bf A}(\alpha(x_1),\ldots,\alpha(x_r))\not =0$,
then, since $\pi_1$ and $\pi_2$ are weakly compatible interpretations, we have that 
$\pi_i(\hat{R}(\alpha(x_1),\ldots, \alpha(x_r)) \not = 0$, for $i=1,2$.
Therefore, since $\pi_1$ and $\pi_2$ are model-defining interpretations, we have that
$\pi_i(\neg \hat{R}(\alpha(x_1),\ldots, \alpha(x_r))  = 0$, for $i=1,2$.  Consequently, we have that
$\pi_i[[\neg\hat{R}(x_1,\ldots,x_r)]](\alpha) = 0$, for $i=1, 2$.

\end{enumerate}

    \item Assume that the formula $\varphi(x_1,\ldots,x_n)$ is of the form $\psi_1(x_1,\ldots,x_n)\lor\psi_2(x_1,\ldots,x_n)$ and that the result holds for the formulas $\psi_1$ and $\psi_2$. Thus,  $$\pi_1[[\psi_i]](\alpha)\not = 0 \quad \text{if and only if}\quad\pi_2[[\psi_i]](\alpha)\not = 0$$ holds, for $i=1,2$. 
    We also have that
    $\pi_i[[\varphi]](\alpha)= \pi_i[[\psi_1]](\alpha)+\pi_i[[\psi_2]](\alpha)$, for $i=1,2$.
Then, the following statements are equivalent:
\begin{enumerate}
\item $\pi_1[[\varphi]](\alpha)\not  = 0$;
\item $\pi_1[[\psi_1]](\alpha)+\pi_1[[\psi_2]](\alpha)\not = 0$;
\item $\pi_1[[\psi_1]](\alpha)\not = 0$ or $\pi_1[[\psi_2]](\alpha)\not = 0$;
\item $\pi_2[[\psi_1]](\alpha)\not = 0$ or $\pi_2[[\psi_2]](\alpha)\not = 0$;
\item $\pi_2[[\varphi]](\alpha)\not  = 0,$
\end{enumerate}
where the equivalence between the third and the fourth statement follows from the induction hypothesis, while the equivalence between the fourth and the fifth statement holds because $\mathbb K$ is zero-sum free.

  \item Assume that the formula $\varphi(x_1,\ldots,x_n)$ is of the form $\psi_1(x_1,\ldots,x_n)\wedge \psi_2(x_1,\ldots,x_n)$ and that the result holds for the formulas $\psi_1$ and $\psi_2$. Thus,  $$\pi_1[[\psi_i]](\alpha)\not = 0 \quad \text{if and only if}\quad\pi_2[[\psi_i]](\alpha)\not = 0$$ holds, for $i=1,2$. 
    We also have that
    $\pi_i[[\varphi]](\alpha)= \pi_i[[\psi_1]](\alpha)\cdot \pi_i[[\psi_2]](\alpha)$, for $i=1,2$.
Then, the following statements are equivalent:
\begin{enumerate}
\item $\pi_1[[\varphi]](\alpha)\not  = 0$;
\item $\pi_1[[\psi_1]](\alpha)\cdot \pi_1[[\psi_2]](\alpha)\not = 0$;
\item $\pi_1[[\psi_1]](\alpha)\not = 0$ and $\pi_1[[\psi_2]](\alpha)\not = 0$;
\item $\pi_2[[\psi_1]](\alpha)\not = 0$ and $\pi_2[[\psi_2]](\alpha)\not = 0$;
\item $\pi_2[[\varphi]](\alpha)\not  = 0,$
\end{enumerate}
where the equivalence between the third and the fourth statement follows from the induction hypothesis, while the equivalence between the fourth and the fifth statement holds because $\mathbb K$ has no zero divisors.

    \item Assume that $\varphi(x_1,\ldots,x_n)$ is
    of the form $\exists x_{n+1}\chi(x_1,\ldots, x_n,x_{n+1})$ and that the result holds for 
    the formula $\chi(x_1,\ldots,x_n,x_{n+1})$. 
    Thus, for all assignments $\alpha$, we have that 
$$ \pi_1[[\chi]](\alpha) \not = 0\quad\text{if and only if}\quad
\pi_2[[\chi]](\alpha) \not = 0.$$
We also have that $\pi_i[[\varphi]](\alpha)= \sum_{b\in A}\pi_i[[\chi]](\alpha[x/b])$, for $i=1,2$. Then the following statements are equivalent:
\begin{enumerate}
\item $\pi_1[[\varphi]](\alpha)\not  = 0$;
\item $\sum_{b\in A}\pi_1[[\chi]](\alpha[x/b])\not = 0$;
\item There is some element $b\in A$ such that  
$\pi_1[[\chi]](\alpha[x/b]) \not = 0$;
\item 
There is some element $b\in A$ such that  
$\pi_2[[\chi]](\alpha[x/b]) \not = 0$;
\item $\sum_{b\in A}\pi_2[[\chi]](\alpha[x/b])\not = 0$;
\item $\pi_2[[\varphi]](\alpha)\not  = 0$,
\end{enumerate}
where the equivalence between the third and the fourth statement follows from the induction hypothesis, while the equivalence between the fourth and the fifth statement holds because $\mathbb K$ is zero-sum free. 

 \item Assume that $\varphi(x_1,\ldots,x_n)$ is
    of the form $\forall x_{n+1}\chi(x_1,\ldots, x_n,x_{n+1})$ and that the result holds for 
    the formula $\chi(x_1,\ldots,x_n,x_{n+1})$. 
    Thus, for all assignments $\alpha$, we have that 
$$ \pi_1[[\chi]](\alpha) \not = 0\quad\text{if and only if}\quad
\pi_2[[\chi]](\alpha) \not = 0.$$
We also have that $\pi_i[[\varphi]](\alpha)= \Pi_{b\in A}\pi_i[[\chi]](\alpha[x/b])$, for $i=1,2$. Then the following statements are equivalent:
\begin{enumerate}
\item $\pi_1[[\varphi]](\alpha)\not  = 0$;
\item $\Pi_{b\in A}\pi_1[[\chi]](\alpha[x/b])\not = 0$;
\item For every element $b\in A$, we have that  
$\pi_1[[\chi]](\alpha[x/b]) \not = 0$;
\item 
For every element $b\in A$, we have that 
$\pi_2[[\chi]](\alpha[x/b]) \not = 0$;
\item $\Pi_{b\in A}\pi_2[[\chi]](\alpha[x/b])\not = 0$;
\item $\pi_2[[\varphi]](\alpha)\not  = 0$,
\end{enumerate}
where the equivalence between the third and the fourth statement follows from the induction hypothesis, while the equivalence between the fourth and the fifth statement holds because $\mathbb K$ has no zero divisors.
\end{itemize}
The proof of the lemma is now complete.
 \end{proof}
 
We now bring the monus query into the picture.
 
\begin{definition}
Let ${\mathbb K}=(K, +, \cdot, 0,1) $ be a positive semiring with monus and let $\tau$ be a relational schema
containing two $n$-ary relation symbols $\hat{R}$ and $\hat{S}$, for some $n\geq 1$.

The \emph{monus query on $\mathbb K$-databases} is
the query  $q_m=\hat{R}\setminus \hat{S}$, i.e., for every    $\mathbb K$-database ${\bf I}$, we have that 
$q_m({\bf I})= R^{\bf I}- S^{\bf I}$, where $-$ is the monus operation on $\mathbb K$.
\end{definition}

If $\mathbb K$ is the Boolean semiring ${\mathbb B}=(\{0,1\}, \vee, \wedge, 0, 1)$, then the monus
query on $\mathbb B$-databases amounts to the difference operation on relations. Hence, it is definable by the $\NNF$-formula $\hat{R}(x)\wedge \neg \hat{S}(x)$
under the canonical-interpretation semantics. As it turns out, the Boolean semiring is the only positive semiring with monus for which this holds.

\begin{theorem} \label{thm:limitations1}
Assume that $\tau$ is a relational schema
containing two unary relation symbols $\hat{R}$ and $\hat{S}$.
Let ${\mathbb K}=(K, +,\cdot, 0, 1)$ be a positive semiring with monus such that $K$ has at least three elements.
 Then the monus query $q_m$ on $\mathbb K$-databases is not definable by any $\NNF$-formula under  canonical-interpretation semantics.
 \end{theorem}

\begin{proof} Since $|K|\geq 3$, there are two non-zero elements $p$ and $q$ in $K$ such that $p\not = q$. We claim that $p-q\not =0$ or $q-p\not = 0$. Indeed, if $p-q=0$
and $p-q=0$, then we would have
$p\preceq q$ and $q\preceq p$, hence we would have $p=q$ because $\preceq$ is a partial order on $K$. So, let us assume, without loss of generality, that $p-q\not = 0$.

Let $a$ be an element in the domain $D$, 
and consider two $\mathbb K$-databases  $\bI=(R^\bI, S^\bI)$ 
and  $\bJ=(R^\bJ, S^\bJ)$ 
with the following properties:

\begin{itemize}
\item $R^\bI(a)=p$ and  $ R^\bJ(a)=p$;
\item $S^\bI(a)=q$ and  $S^\bJ(a)=p$;
\item  $R^\bI(b)= R^\bJ(b)=S^\bI(b)= 
S^\bJ(b)= 0$, for all other elements $b$ in $D$.
\end{itemize}

Consider
 the relational algebra expression $\hat{R}\setminus \hat{S}$. 
Clearly,  
$$(\hat{R}\setminus \hat{S})^\bI(a)=R^\bI(a)- S^\bI(a) =p-q \not = 0\quad\text{and}\quad 
(\hat{R}\setminus \hat{S})^\bJ(a)=R^\bJ(a)- S^\bJ(a) =p-p = 0.$$  

Furthermore, we have that the following hold:
\begin{itemize}
\item 
${\bf A}(\bI)=(\{a\},R^\bI, S^\bI)$
and ${\bf A}(\bJ)=(\{a\},R^\bJ, S^\bJ)$;
\item  the canonical interpretation $\pdi$  is the function on the literals from $\{a\}$ such that $\pdi(\hat{R}(a))=R^\bI(a)=p$, $\pdi (\neg \hat{R}(a))=0$, $\pdi(\hat{S}(a))=S^\bI(a)=q$, and $\pdi(\neg \hat{S}(a))=0$.

\item  the canonical interpretation $\pdij$  is the function on the literals from $\{a\}$ such that $\pdij(\hat{R}(a))=R^\bJ(a)=p$, $\pdij (\neg \hat{R}(a))=0$, $\pdij(\hat{S}(a))=S^\bJ(a)=p$, and $\pdij(\neg \hat{S}(a))=0$.
\end{itemize}
Towards a contradiction, assume that there is a $\NNF$-formula $\varphi(x_1)$ that defines the monus query $q_m$ on $\mathbb K$-databases.  Consider an assignment $\alpha$ such that $\alpha(x_1)=a$.
Then we must have that
$$(\hat{R}\setminus \hat{S})^\bI(a)= p-q = \pdi[[\varphi]](\alpha)\quad\text{and}\quad(\hat{R}\setminus \hat{S})^\bJ(a)=0=\pdij[[\varphi]](\alpha).$$
In particular and since $p-q \not = 0$, we must have that $\pdi[[\varphi]](\alpha)\not = 0$, while $\pdij[[\varphi]](\alpha)=0$.
However, the interpretation $\pdij$ is weakly compatible with the structure ${\bf A}(\bI)$. Hence, Lemma \ref{lem:key} implies that $\pdi[[\varphi]](\alpha)\not = 0$ if and only if $\pdij[[\varphi]](\alpha)\not =0$, which is a contradiction.
Thus, no $\NNF$-formula can define the monus query $q_m$ on $\mathbb K$-databases.
\end{proof}

It is  instructive to pinpoint where the proof of Theorem \ref{thm:limitations1} breaks down if $\mathbb K$ is the Boolean semiring $\mathbb B$ and  we take $p=1$ and $q=0$ as our two different elements. When one attempts to carry the proof over to $\mathbb B$, then the databases $\bI$ and $\bJ$ are as follows:
\begin{itemize}
\item $\bI=(R^\bI, S^\bI)$, where
$R^\bI=\{a\}$ and $S^\bI= \emptyset$;
\item $\bJ=(R^\bJ, S^\bJ)$, where
$R^\bJ=\{a\}$ and $S^\bJ= \{a\}$.
\end{itemize}
In this case, however, the interpretation $\pdij$ is \emph{not} weakly compatible with ${\bf A}(\bI)$; thus, Lemma \ref{lem:key} cannot be applied and the proof  does not go through.

The preceding Theorem \ref{thm:limitations1} reveals a striking limitation in the expressive power of the canonical-interpretation semantics.  Moreover, the proof of this theorem actually shows that the same limitation holds true for the compatible-interpretation semantics. The reason is that the $\mathbb K$-databases used in the proof have singleton sets as their active domains and each of  their relations has non-empty support. If $\bf I$ is such a database,  then the \emph{only} compatible interpretation on $\adom(\bI)$ is the canonical one. Therefore, the proof of Theorem \ref{thm:limitations1} establishes the following stronger result.

\begin{theorem} \label{thm:limitations2}
Assume that $\tau$ is a relational schema
containing two unary relation symbols $\hat{R}$ and $\hat{S}$.
Let ${\mathbb K}=(K, +,\cdot, 0, 1)$ be a positive semiring with monus such that $K$ has at least three elements.
 Then the monus query $q_m$ on $\mathbb K$-databases is not definable by any $\NNF$-formula under  compatible-interpretation semantics.

 \end{theorem}

 Theorem \ref{thm:limitations2} implies that there is \emph{no}  version of Codd's theorem involving relational algebra with monus on the one side and compatible-based interpretations on the other side. In particular, since the monus operation on bags is used in the semantics of SQL, we have the following implication for SQL.

 \begin{corollary} \label{corr2}
The semantics of SQL cannot be captured using  compatible-interpretation semantics; in particular, they cannot be captured using canonical-interpretation semantics.
 \end{corollary}

Recall that universal quantification is not allowed in  the basic relational calculus $\mathcal{BRC}$. In \cite{DBLP:journals/pacmmod/BadiaKN25}, relational calculus $\mathcal{RC}$ was defined as the extension of $\mathcal{BRC}$ with universal quantification. It is not hard to see that $\mathcal{RC}$ properly subsumes $\NNF$ under canonical-interpretation semantics.
To see this, let $\varphi$ be an arbitrary $\NNF$-formula.  We claim that there exists a $\mathcal{RC}$-formula $\psi$ such that $\pi_{{\bf A}(\bI)}[[\varphi]] = 
\psi^{{\bf A}(\bI)}$, for every $\mathbb K$-database $\bI$.   For this, all we have to do is to replace every occurrence of a negated atom $\neg \hat{R}(x_1,\ldots,x_n)$ by the $\mathcal{BRC}$-formula $(x_1=x_1) \butnot \supp (R(x_1,\ldots,x_n))$. This subsumption is a proper one, because, by Theorem \ref{thm:limitations1}, the monus query is not definable by any $\NNF$-formula under canonical-interpretation semantics.
In \cite{DBLP:journals/pacmmod/BadiaKN25}, it was shown that relational calculus $\mathcal{RC} $ has the same expressive power as relational algebra $\mathcal{RA}$, which is the extension of basic relational algebra $\mathcal{BRA}$ with the division operation. Consequently, $\mathcal{RA}$ properly subsumes $\NNF$ under canonical-interpretation semantics.

\bibliographystyle{alpha}
\bibliography{semirings}

\end{document}